\documentclass[runningheads,a4paper]{llncs}

\usepackage{mathrsfs}
\usepackage{url}

\usepackage{amssymb}
\usepackage{stfloats}

\usepackage{graphicx}

\usepackage{psfig}
\usepackage{bm}

\usepackage{epstopdf}

\usepackage{times}
\usepackage{helvet}
\usepackage{courier}
\usepackage{url}

\usepackage{float}
\usepackage{subfigure}
\usepackage{subfig}
\usepackage{caption}

\usepackage{url}
\usepackage{booktabs}
\usepackage{multirow}

\usepackage{overpic}

\usepackage{CJK}
 \usepackage{amsmath}
 
\begin{document}
%
\title{Multi-view shaker detection: Insights from a noise-immune influence analysis Perspective}

\titlerunning{Short-term shaker detection }

\mainmatter  

%

%
%
\author{Chang Liao}
%
\institute{Dongguan Securities Company Limited,  China }
%
%

\maketitle

\begin{abstract}
Entities whose changes will significantly affect others in a networked system are called shakers.  In recent years, some models have been proposed to detect such shaker from evolving entities.  However, limited work has focused on shaker detection in very short term, which has many real-world applications. For example,  in financial market, it can enable both investors and governors to quickly respond to rapid changes. Under the short-term setting, conventional methods may suffer from limited data sample problems and are sensitive to cynical manipulations, leading to unreliable results. Fortunately, there are  multi-attribute evolution records available, which  can provide compatible and complementary information.  In this paper,   we investigate how to learn reliable  influence results from the short-term multi-attribute evolution records. We  call  entities with consistent influence among different views in short term as multi-view  shakers and study the new problem of multi-view shaker detection.
  We identify the challenges as follows: (1) how to jointly detect short-term shakers and model conflicting influence results among different views? (2) how to filter spurious influence relation in each individual view for robust influence inference? In response, a novel solution, called Robust Influence Network ($\mathbf{RIN}$) from a noise-immune influence analysis perspective is proposed, where  the possible outliers are well modelled jointly with multi-view shaker detection task.   More specifically, we  learn the influence relation from each view and  transform influence relation from different views into an intermediate representation. In the meantime, we uncover both the inconsistent and spurious outliers.
 The experimental results demonstrate the effectiveness of our approach.
\end{abstract}

\section{Introduction}

\vspace{-0.232cm}

In real applications, it is well recognized that there are some influential entities whose evolution can have a significant impact on the dynamics of others. As with \cite{shi2011discovering1,shi2013dynamic}, entities with such high impact are shakers. For example, there is considerable evidence pointing to existence of the lead-lag effect (influence relation) in stock market, where some firms' stock prices show a delayed reaction to price movements of other firms (\cite{Hou2007Industry}). An ability to track such influence relation within a dynamically evolving system not only probes
our understanding of complex systems, but also has important
implications in a wide range of domains (\cite{Lozano2010Spatial,Liu2011Discovering1}). These influential entities can act as leading indicators, which are useful for alerting people and organizations to increase response readiness towards critical situations. This is especially important in financial market, when entities are termed  as currencies, foreign exchanges or equities. The application cases are motivated as follows: (1) extraction of risk entities ($\textit{e.g.}$, the 2008 $Lehman$ $Brothers$ $bankruptcy$ leading to the world financial crisis, and $Portugal/$ $Greek$ $sovereign$ $debt$ $risks$ causing the global financial market turmoil) can help market governors prevent system risks in advance; (2) distinguishing upstream and downstream industries ($\textit{e.g.}$, the flourishing of iron, cement and machinery domain bringing in the prosperity of real estate, consumption, culture industries) can help market investors make better investment strategies.

\begin{figure*}[t]
\centering

\includegraphics [width=0.9898278048902648908949671998287488\textwidth,height=0.287589754730124228323453846502762346840134\textwidth]{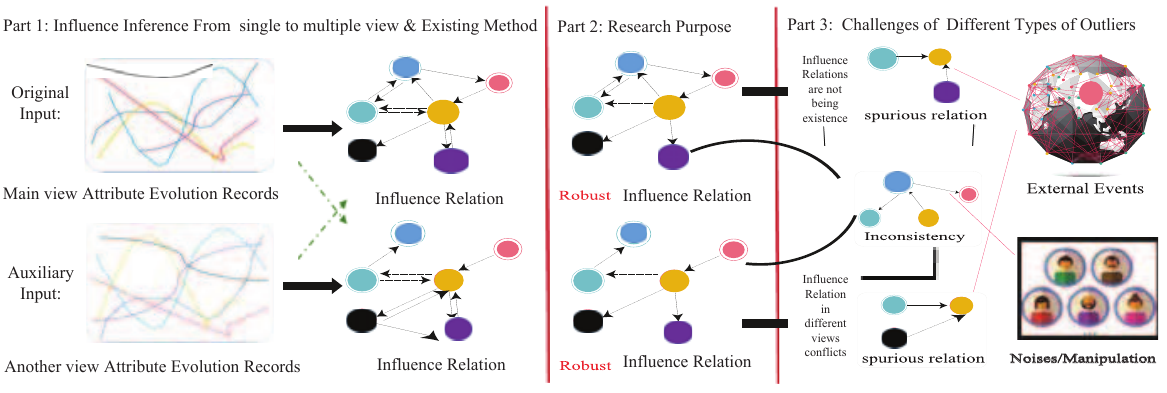}

\vspace{-0.2cm}

\caption{An overview (example) of Multi-view Influence Inference. Part 1 shows the influence inference from time series by existing method. Part 2 shows the research purpose - robust influence inference. Part 3 shows the Potential Outliers. More specific, there are 6 entities with 2 views. By modeling conflicting influence relation results brought by noise/ manipulations as well as spurious influence relation results caused by external events, we can achieve a robust influence relation for each view (Balls represent the entities and edges represent influence relation strength).  }
\vspace{-0.346531cm}

\end{figure*}

In recent years, there is some work proposed to infer influence relation from time series data  on a single view of the entities, as shown in the first part Figure 1. \cite{shi2011discovering1} proposes to use ``cascading graph'' to detect shakers. \cite{shi2013dynamic}
proposes to model evolving influence relation between entities on long-time series data. However, the influence relation between entities in financial market is fast changing and developing after the 2008 economic crisis, where only short-term influence relation counts. However, it might suffer from limited data samples and possible manipulation problems  when directly learnt on very short time period. To be more specific, they are increasingly faced with problems where the number of variables or features (such as the amount of stocks) is larger than the number of observations (such as stock price series length). What's worse,  there are some noises or even manipulation crimes by intentionally affecting or controlling the movement of certain entities among short-time intervals, which occurs frequently in financial market. Under such settings,  the model without accounting for possible manipulations under limited data samples  may provide potential conflicting  influence relation results. Besides stock market, this setting can also accommodate a variety of potential scenarios, such as influence inference between  new products in online E-commerce, especially for electronic prod, etc.

While multiple
view data provides compatible and complementary information, it
becomes natural to integrate it together to obtain better results.  For instance, stock price information and trading volume information are complementary - useful for distinguishing influence relation in different ways. It is then natural to integrate them together to obtain better performance rather than relying on a single view. The assumption is that  influence relation across different views should be consistent, which can act as a constraint for solving limited data sample issues.    And if there are controversial influence results between two entities across different views, the entities might be possibly manipulated. To fill this gap of the lack of short-term shaker detection  and motivated by the benefits of multiple view data, we  formulate it as a new problem - multiple view shaker detection. Despite of the huge benefits in multiple view information, it does not work well without properly modelled, as  shown in the third part of  Figure 1.  We describe and address the following two challenging problems:

On the one hand, the problem lies in how to exploit the learnt influence relation from different views and consider potential outliers simultaneously for influence relationship inference.  For instance, the influence relation learnt from stock price view is significantly different from that learnt from stock volume view, presenting expanded scale of influence relation strength. Direct incorporation of them is harmful for the results. Meanwhile, multi-view inconsistencies (outliers) exist when
conflicting influence relations are inferred between entities from different views.  It is due to possible manipulation occurs on the given short time interval. Simply pooling all views together into a shared latent space may degrade the overall performance.

On the other hand, the external events on certain entities can contribute to their activity evolutions.  For example, a great many stock prices fluctuate with currency market, commodity market, $\textit{etc.}$  And straightforwardly, it would result in the appearance of spurious influence relations between arbitrary entities due to the lack of considerations of these unobserved events. These outliers are consistent from multiple view perspectives and seem normal, which is then largely neglected in multi-view settings. Consequently, the influence relation strength inference results from multiple time series data are dramatically harmed when ignoring them.

Since there are no previous methods that are directly applicable  to track the above challenges, we propose a multi-objective optimization framework to integrate influence inference and outlier detection tasks together. As an example of multiple view shaker detection  in Figure 1, we attempt to recover robust influence relation from multiple time series data in short time interval, where conflicting relation outlier and spurious relation outlier exist. By representing influence relation in both
latent space and original space, our model can jointly model and discover these two types of outliers, enabling to infer more robust influence relations. Finally, multi-view shakers are detected based on information diffusion modeling. The  contributions are  as follows:
\itemindent -22pt

\vspace{-0.2153642cm}

\begin{itemize}
\item  We motivate to formulate a new problem - multiple view shakers detection from time series, which is immensely helpful in many real applications, especially in financial market. Moreover, we propose to infer robust influence relation together with identifying and removing outliers.  To the best of our knowledge, the problem of influence analysis has not been studied in this context so far.

    \vspace{-0.0512cm}

\item A robust influence inference method ($RIN$) that integrates multi-objective optimization with low rank and group sparse constraints is proposed.  The problem is solved by Augmented Lagrange Multiplier ($ALM$) algorithm. We  conduct extensive experiments to demonstrate the effectiveness of our method, apply it to  stock market data and derive new metrics to measure the results.
\end{itemize}

\vspace{-0.7542cm}

\section{Problem Statement}
\vspace{-0.32167852cm}
In this section, we introduce several concepts and then formulate the problem. Table 1 shows the important notations we will use in this paper.
\vspace{-0.0107852cm}

\textbf{Multi-view (Multi-attribute) Evolving Entities:} Multi-view (Multi-attribute\footnote{Attributes that are evolving and continuous are selected, not meaning all attributes are.}) Evolving Entities are defined as entities with multi-attribute evolution records   $\mathcal{Z} = [ {Z^1},{Z^{2}},...,{Z^V}]  \in {R^{L \times D}}$ with continuous value. The records are   with $|N|$ entities and $V$ views, where  $D=|N|\times V$ and $L$ is time interval. As an example in Figure 2(a), the multi-attribute evolution record of an entity is represented as multiple time series. Among it, each row represents the movement value for each entity in each view (attribute) at a particular time, while each column indicates the movement evolution time series for a particular entity in a particular view.

\begin{table}
\caption{Several important mathematical notations}

\centering
\begin{tabular}{c|c|c|c}
\toprule
 Notations & Meaning & Notations & Meaning\\
\midrule
       $\mathcal{Z}$, $\mathcal{X}$, $\mathcal{Y}$ &   Multiple Time Series  &  $\varPhi$ & Inconsistent Relation \\

         $W$,$\hat{W}$,$\tilde{W}$ &  Coarse/Biased/Robust  Influence Matrix   &  $\varTheta$ &  Spurious Relations      \\

    ${\rho _1}$, ${\rho _1}$, ${\rho _3}$ & Regularization Parameters       & $k$ & Subspace Matrix Rank  \\

          $u$ &  Lagrange Weight   & $r$ & Influence Time Duration      \\

        $A$ & Shared Feature Space &  $f$ & Entity Influence Strength  \\

       $E$ &  Combined Influence Matrix  & $v$, $S(L)$  &    View Indicator/ Instance Size \\

        ${\lambda _1}$, ${\lambda _2}$, ${\lambda _3}$ & Lagrange Multipliers Parameters & $\bar{Wr}$, $Wr$  & Synthetic Experiment Parameters\\

        $p_1$, $p_2$, $wp$, $q$ & Algorithmic Trading Parameters &  $U$, $\mathbf{m}$ & Market Simulation Parameters \\

\bottomrule
\end{tabular}

\end{table}

\textbf{Influence Network:}  An entity is thought to  affect another when its activity time series unveils lead-lag relations with the other.  In our settings, it is assumed that the activity of an entity  can be influenced by the other entities under the \textbf{Markovian assumption.}, where influence network can be constructed.  As in Figure 2(b), influence matrix $\hat{W} \in {R^{|N| \times |N|}}$ represents the real-valued influence relation strength between entities. Each view has an influence matrix, where the main view is the view (attribute) we are interested in, other views can be taken as auxiliary information.

Compared with Granger influence (\cite{granger1969investigating}), there are mainly two differences: (1)  the activity of an entity at $t$ only depends on others at $t-1$ and influence network is instantaneous, where the time before $t-1$ is not considered; (2) a combination of influence relation is taken into account as opposed to every two   time series pairs.

\textbf{Outlier \uppercase\expandafter{\romannumeral1}} $\varPhi \in {R^{|N| \times |N|}}$ is defined as inconsistent influence relation across different views. \textbf{Outlier \uppercase\expandafter{\romannumeral2}} $\varTheta \in {R^{|N| \times |N|}}$ is defined as spurious relations obtained due to the
latent  events. A typical example about the two types of outlier is shown in Figure 2(c).

Obviously, \textbf{Outlier \uppercase\expandafter{\romannumeral1}} models the case where noises or any possibly manipulation crimes comes into being. When abnormal shock occurs, the learnt influence relation is possibly manipulated.  To be concrete, though the impact that it affects others is still accurate, the influence relation strength  that others have on it is biased. \textbf{Outlier \uppercase\expandafter{\romannumeral2}} would model the external events. In the stock market example,  like policies or currencies, that affect the stock prices. While we aim to compare the shaker's ability to predict the whole market change, adding other information  might improve the result. Here, the spurious outlier can represent and model such  ``relevant'' information.

\begin{figure}[t]
\vspace{-0.2cm}
\centering
\subfigure[Multi-View Evolution Records]{\includegraphics [width=0.23245402340983584758\textwidth,height=0.1752783120286851243048402363064022147\textwidth]{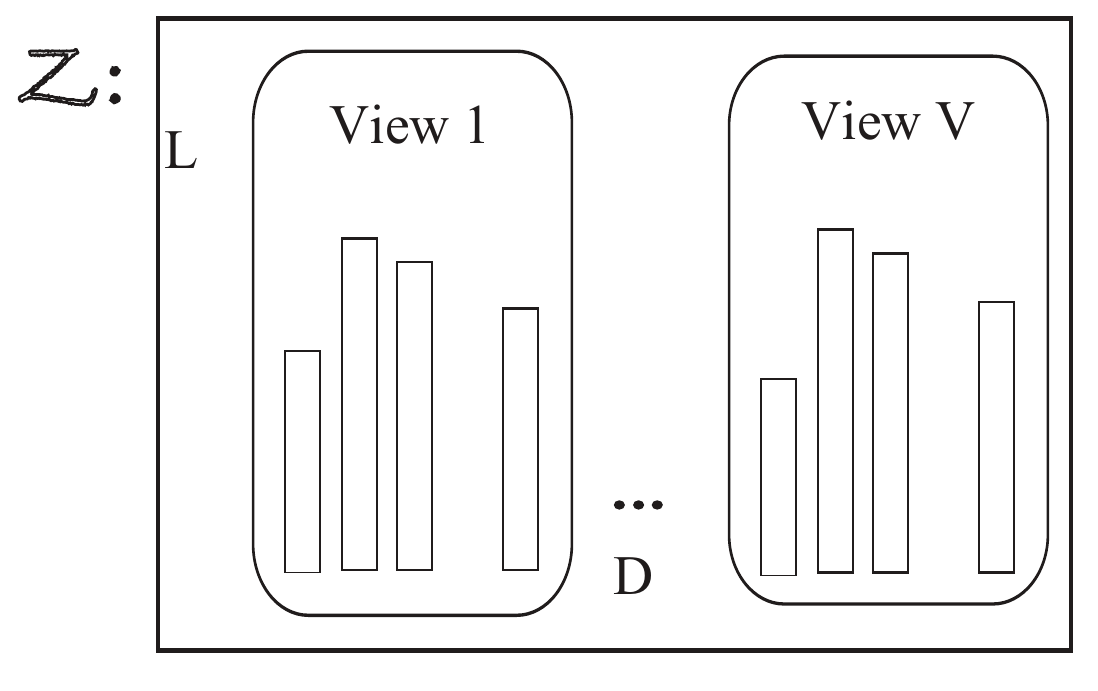}}
\subfigure[Multi-View Influence Matrix]{\includegraphics [width=0.21972032402340983584758\textwidth,height=0.175278312028685230468402363064022147\textwidth]{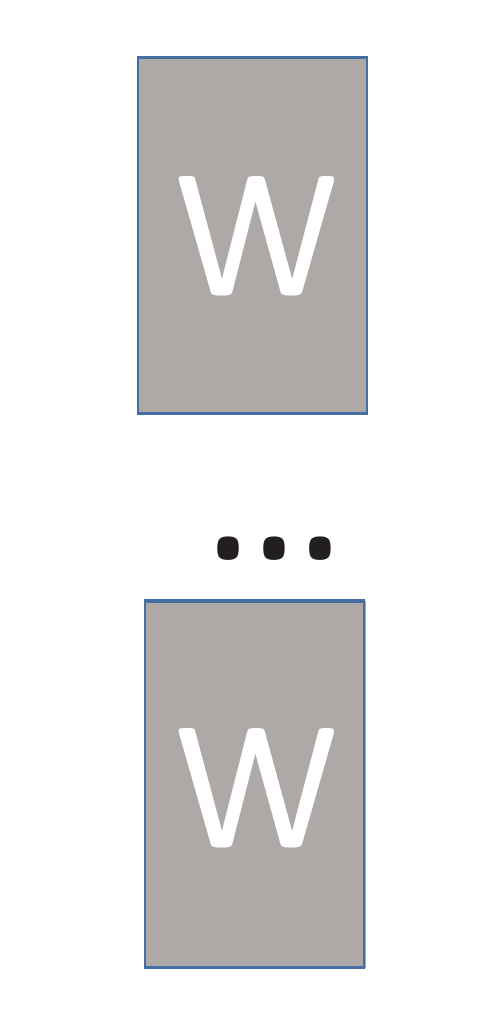}}
\subfigure[Two Types of Outliers]{\includegraphics [width=0.18202453245402340983584758\textwidth,height=0.1752783120286851243048402363064022147\textwidth]{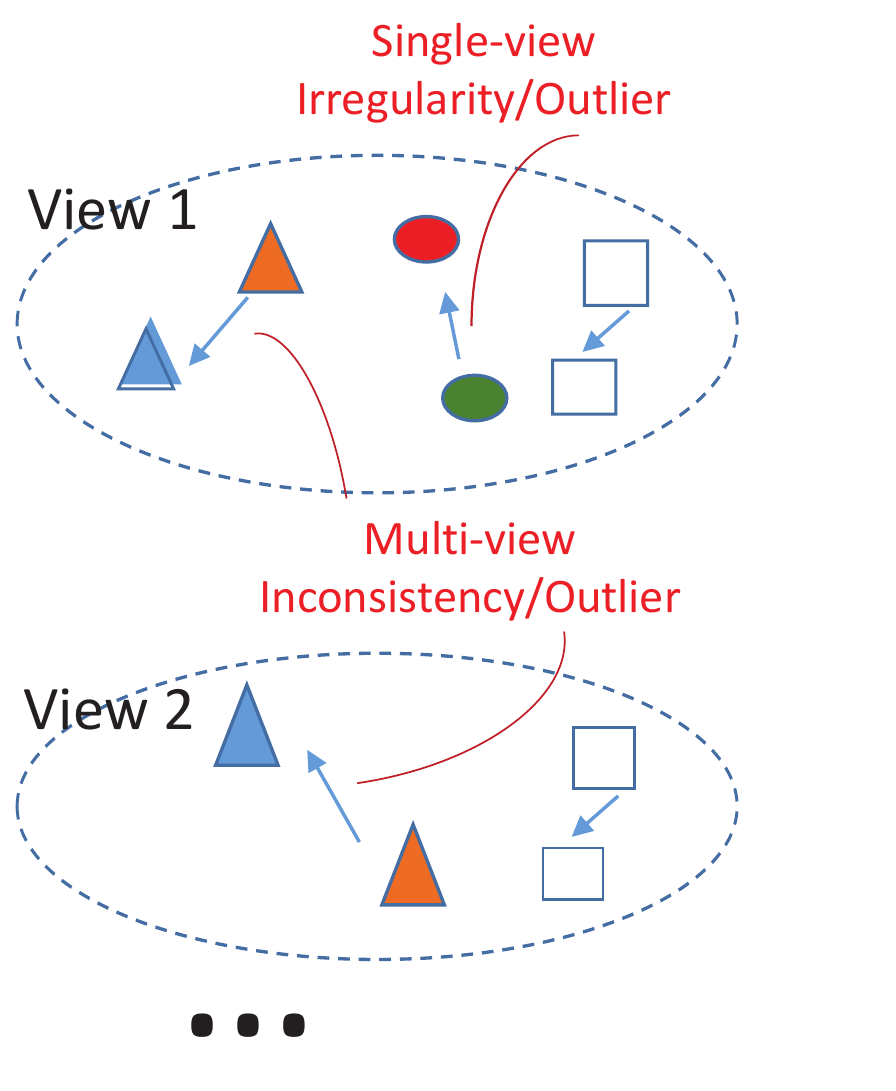}}
\subfigure[Intuitive: From Single to Multiple Views]{\includegraphics [width=0.26432402340983584758\textwidth,height=0.175278312028685230468402363064022147\textwidth]{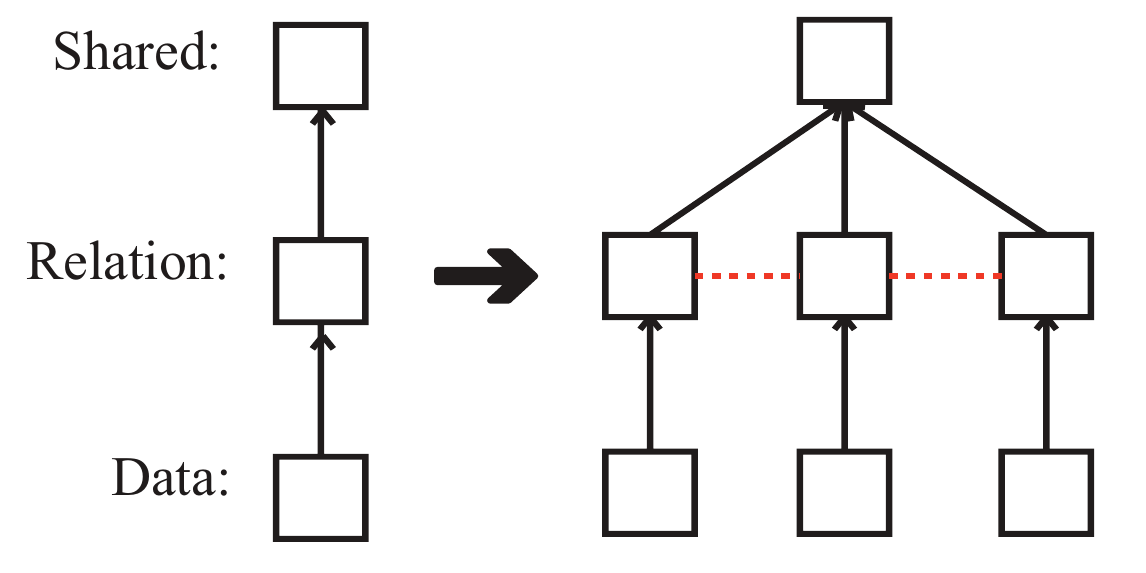}}

\vspace{-0.1382cm}
\caption{Terminology Illustrations }
\label{fig:5}

\vspace{-0.32cm}
\end{figure}

\textbf{Multiple View Shakers Detection:} Define multiple view shaker(s) as an entity (entities) whose evolution $X_t^v(i)$ of main view (attribute) $v$ at time $t$ can lead to maximal changes subsequently in the whole system. We claim multi-view shaker is a special kind of shakers (\cite{shi2011discovering1}), which focuses on very short term and is insensitive to different types of outliers.

By \underline{r}obust \underline{i}nfluence \underline{n}etwork  construction from multi-view (multi-attribute) evolving entities, a new algorithm called $RIN$ is proposed. It aims at exploiting information from single view time series to multiple view time series, as shown in Figure 2(d).  The model \textbf{Input:} (1) multi-attribute evolution time series $\mathcal{X} = [ {X^1},{X^{2}},...,{X^V}]  \in {R^{S \times D}}$ from the beginning time $t_0$ to the end time $t_{S-1}$ ; (2) The following/consecutive time series for corresponding entities with the same time interval  as  $\mathcal{Y} = [ {Y^1},{Y^2}...,{Y^V}]  \in {R^{S \times D}}$ from the beginning time $t_1$ to the end time $t_{S}$.  The model \textbf{Output:} The influence relation matrix $\hat{W}\in {R^{|N| \times |N|}}$ can be estimated together with removing Outlier \uppercase\expandafter{\romannumeral1}  and Outlier \uppercase\expandafter{\romannumeral2}. And moreover, multi-view shakers are  detected by what-if analysis over the learnt influence network $\tilde{W}$.

\vspace{-0.15cm}

\section{Influence Mining Model: RIN}

\vspace{-0.135cm}
%

The proposed $RIN$ is a two-stage approach, as is shown in Figure 3. In the first phase, we  learn the influence relation strength with multiple views. More importantly, we argue that by  mapping influence relation from each view in latent space, our model can simultaneously infer influence relation and filter outliers in a more robust way by low rank and group sparse constraints. In the second phase, multiple view shakers are detected based on influence diffusion modeling.

\begin{figure}[h]
\centering

\vspace{-0.243cm}

\includegraphics [width=0.9528797042686898986895\textwidth,height=0.4863528502830264679872583075096234567698535861049832\textwidth]{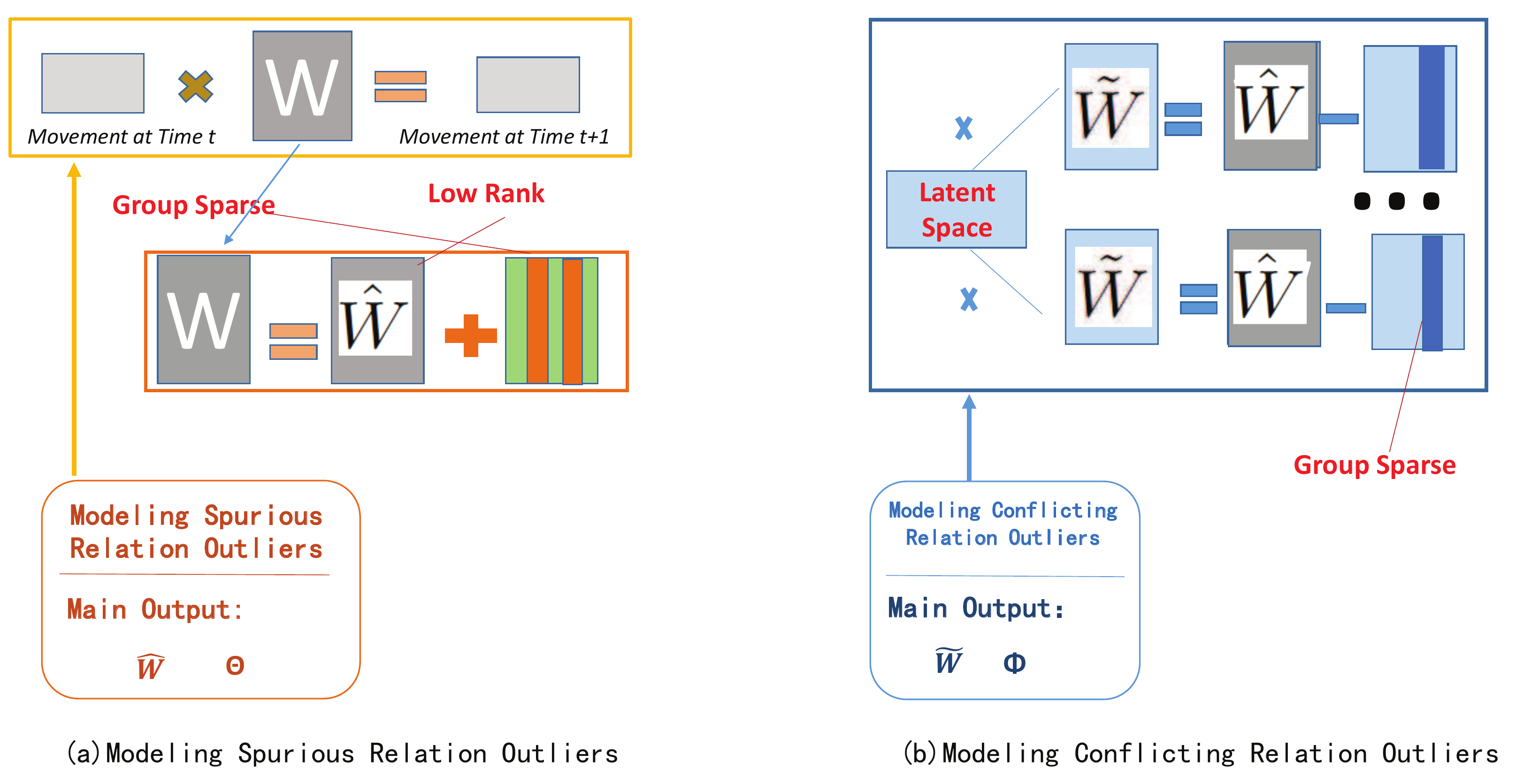}

\vspace{-0.081243cm}
\caption{Overview of the proposed method ($RIN$), which consists of influence network construction  and multi-view shaker detection. To obtain robust influence relation, multi-attribute (multi-view) evolution information is utilized and two types of outliers are considered: (1) Low rank and group sparse constraints are incorporated to detect spurious influence relation outliers; (2) Group sparse constraints are added  to
model inconsistent influence relation across views in the shared latent space. }

\vspace{-0.43cm}
\end{figure}

\vspace{-0.321862cm}

\subsection{Consistent Spurious Relation Outliers Modeling}
\vspace{-0.2cm}

As for influence inference, we
aim to minimise the discrepancy between the ground truth $Y_{S+1}$ and our prediction ${X_S}{W}$.
As shown in the Phase 1.1 in Figure 3, given attribute evolution  time series value $X_S$ at time $[t_0,t_S)$  and consecutive  series $Y_{S+1}$ at time $[t_1,t_{S+1})$ for a set of  entities from each view, the $RIN$ assumes  the movement of entities at time $t+1$ is affected by the movement of their surroundings at time $t$, which is denoted as ${Y_{S + 1}} = {X_S}W$,  under the first-order Markovian assumption. Taking the \textbf{Consistent spurious relations} generated by the external event further,  we denote  the common spurious influence relation scores as $\varTheta \in {R^{N \times N}}$.  As shown in the Phase 1.3 in Figure 3,   we model the robust influence relation together with consistent spurious relations as follows:

\begin{small}
\begin{equation}
\vspace{-0.12cm}
\begin{array}{l}
 \mathop {\min }\limits_{{W^v},{{\hat W}^v},{\Theta ^v}} \sum\nolimits_v {||Y_{S + 1}^v - X_S^v{W^v}||_F^2 + } {\rho _1}||{{\hat W}^v}|{|_*} + {\rho _2}||{\Theta ^v}|{|_{2,1}} \\
  \\
 s.t.{\rm{   }}{W^v} = {{\hat W}^v} + {\Theta ^v} \\
 \end{array}
\end{equation}
\end{small}

where  the ${\ell_{2,1}}$ - norm regularization term in matrix $\varTheta$ induces the desirable group-sparse structures   for identifying the outliers (spurious relations). The rationale is that some entities are heavily affected by external events, and the corresponding columns for these ones in $\varTheta$ are non-zero compared with others. The weighing parameter $\rho_1$ of nuclear norm is to force the learnt influence relation matrix low rank, and weighing parameter $\rho_2$ is used in distinguishing inconsistent outlier. By decomposing $W^v$ into influence matrix $\hat{W}^v$ and consistent spurious relation matrix $\varTheta^v$, influence matrix $\hat{W}^v$ becomes more robust to external events compared with  $\tilde{W}$. The low rank constraints in the influence matrix ${\hat{W}}^v$ preserves the
relatedness structure and provides time series noises tolerant properties.


The problem in Equation (1) is a mixed nuclear norm and $\ell_{2,1}$ norm optimization with orthogonality constraints. As with these settings, the problem cannot be solved directly with existing standard optimization tools. Inspired by Augmented Lagrange Multiplier ($ALM$) algorithm (\cite{lin2010augmented}), the basic idea is to  optimize the variables independently in an iterative manner, where $\lambda$ are casted as the Lagrange Multipliers as follows:

\vspace{-0.2cm}

\begin{small}
\begin{equation}
\begin{array}{l}
 \mathop {\min }\limits_{{W^v},{{\hat W}^v},{\Theta ^v}} \sum\nolimits_v {||Y_{S + 1}^v - X_S^v{W^v}||_F^2 + } {\rho _1}||{{\hat W}^v}|{|_*} + {\rho _2}||{\Theta ^v}|{|_{2,1}} \\
  \\
  + \sum\nolimits_v {u||{W^v} - {{\hat W}^v} - {\Theta ^v}||_F^2 +  < {\lambda _1},{W^v} - {{\hat W}^v} - {\Theta ^v} > }  \\
 \end{array}
\end{equation}
\end{small}

\vspace{-0.2cm}
The iteration procedure to update the parameters is carried on until converges. Specifically, the variables $W$,  $\varTheta$, $A$ and $\hat{W}$ are updated as follows, where $W_{(i)}$,  $\varTheta_{(i)}$ and  $\hat{W_{(i)}}$ are the values at $i$-th  iteration respectively.

\textbf{Update $W_{(i)}$:} Fix $\varPhi_{(i)}$ and $\tilde{W}_{(i)}$, the function for $W_{(i)}$ is as:

\vspace{-0.12cm}
\begin{equation}
\begin{array}{l}
\mathop {\min }\limits_{{W^v}} \sum\nolimits_v {||Y_{S + 1}^v - X_S^v{W^v}||_F^2 + } u||{W^v} - {\hat W^v} - {\Theta ^v}||_F^2 +  < {\lambda _1},{W^v} - {\hat W^v} - {\Theta ^v} >
 \end{array}
\end{equation}

\vspace{-0.12cm}

We can solve it by gradient descent.

\textbf{Update $\varTheta_{(i)}$:} Fix  $\tilde{W}_{(i)}$ and $\hat{W}_{(i)}$, the function with respect to $\varTheta_{(i)}$ turns to be:

\vspace{-0.3cm}
\begin{equation}
\begin{array}{l}

\mathop {\min }\limits_{{\Theta ^v}} \sum\nolimits_v {u||{W^v} - {{\hat W}^v} - {\Theta ^v}||_F^2 +  < {\lambda _1},{W^v} - {{\hat W}^v} - {\Theta ^v} > }  + {\rho _2}||{\Theta ^v}|{|_{2,1}}

 \end{array}
\end{equation}

\vspace{-0.3cm}

Each column vector of $\varTheta_{(i)}$ can be optimized separately\cite{liu2009multi}.

\textbf{Update $\hat{W}_{(i)}$:} Fix $\tilde{W}_{(i)}$ and $\varTheta_{(i)}$, the function with respect to $W_{(i)}$ is as:

\vspace{-0.3cm}
\begin{equation}
\begin{array}{l}
\mathop {\min }\limits_{{{\hat W}^v}} \sum\nolimits_v {u||{W^v} - {{\hat W}^v} - {\Theta ^v}||_F^2 +  < {\lambda _1},{W^v} - {{\hat W}^v} - {\Theta ^v} > }  + {\rho _1}||{\hat W^v}|{|_*}

 \end{array}
\end{equation}

\vspace{-0.3cm}

 The derivative cannot
be easily calculated due to trace norm term $||\hat{W}_{(i)}^v|{|_*}$.
Hence, we adopt a novel approach
using the singular value decomposition techniques and
then applying soft-thresholding on the singular values. This computation procedure above is equal to matrix shrinkage operator (\cite{cai2010singular}), and we can just solve it through singular value decomposition (SVD).

\vspace{-0.1385862cm}

\subsection{Influence Network with Multiple Views}

\vspace{-0.1862cm}

Apart from the main view, other views can be obtained as augmented information to improve the inference performance.   That is, influence relations learnt from different views should be made similar by projecting into a shared latent space as ${A^v}^T{\hat{W}^v} = {A^{v'}}^T{\hat{W}^{v'}}$. In other words, the influence relations between the same entities should  be highly similar among different views. When considering conflicting relation among different views problem, we denote influence relation matrix as $\tilde{W} \in {R^{N \times N}}$, the multi-view inconsistent influence relationship strength score for each view as  $\varPhi \in {R^{N \times N}}$. More specifically,  the influence relation matrix $\tilde{W} \in {R^{N \times N}}$ can be learnt by solving

 \vspace{-0.12cm}
\begin{small}
\begin{equation}
\vspace{-0.12cm}
\begin{array}{l}
 \mathop {\min }\limits_{{A^v}{{\tilde W}^v},{\Phi ^v}} \sum\nolimits_v {\sum\nolimits_{v'} {||{A^v}^T{{\tilde W}^v} - {A^{v'}}^T{{\tilde W}^{v'}}||_F^2 + {\rho _3}} ||{\Phi ^v}|{|_{2,1}}}  \\
  \\
 s.t.{\rm{    }}{A^{\rm{v}}}^T{A^v} = I,{{\hat W}^v} = {{\tilde W}^v} + {\Phi ^v} \\
 \end{array}
\end{equation}
\end{small}

where $\lambda$ is a weighing parameter about the relative importance of the corresponding objective. The projection matrix $A^T$  is in a low dimensional space, meaning the different view influence relations preserve the principal coupled patterns in the lower dimensional space for each entity. $A^TA=I$ denotes an orthogonal constraint which makes the projection matrix $A^T$ unique for each view. Orthogonality means that any two basis vectors are orthogonal to each other, ensuring compactness for projecting influence relations.

We assume that $\hat{W}^v$ comprises of influence relation $\tilde{W}^v$ and conflicting relation ${\varPhi^{v}}$.
While there is not enough information to perfectly distinguish outliers and normal parts, we assume outliers are in small quantity compared with normal parts, where  $\ell_{2,1}$-norm in ${\varPhi^{v}}$ is to keep only few composites of the inconsistency score non-zero. For instance, while stock manipulators can increase the value of some attribute (view) intentionally, they can hardly handle a great many. In other words, we assume some vectors in $W^v$ are corrupted by manipulation/ crimes, while others are clean. In such way, we can make the influence relation $\tilde{W}^v$ insensitive to view conflicting outliers as opposed to $\hat{W}^v$, where ${\rho_3}$ is a weighting parameter.

Intuitively, faced with noises or manipulation situations, a straightforward way is to first discover such outlier behaviors, and then conduct influence inference task. However, it is widely accepted that  many influence relations can be discovered when there're abnormal shocks. Motivated by it, we are here not to discard it from entity time series data, but attempt to handle it well. This is  why we consider and filter such noisy influence relations on relationship perspectives, rather than on entity perspectives.

The problem in Equation (6) is a mixed nuclear norm and $\ell_{2,1}$ norm optimization with orthogonality constraints. As with these settings, the problem cannot be solved directly with existing standard optimization tools. Inspired by Augmented Lagrange Multiplier ($ALM$) algorithm (\cite{lin2010augmented}), the basic idea is to  optimize the variables independently in an iterative manner, where $\lambda$ are casted as the Lagrange Multipliers as follows:

\vspace{-0.2cm}

\begin{small}
\begin{equation}
\begin{array}{l}
 \mathop {\min }\limits_{{A^v}{{\tilde W}^v},{\Phi ^v}} {\rho _3}||{\Phi ^v}|{|_{2,1}} + \sum\nolimits_v {\sum\nolimits_{v'} {||{A^v}^T{{\tilde W}^v} - {A^{v'}}^T{{\tilde W}^{v'}}||_F^2} }  \\
  \\
  + \sum\nolimits_v {u||{{\hat W}^v} - {{\tilde W}^v} - {\Phi ^v}||_F^2 +  < {\lambda _2},{{\hat W}^v} - {{\tilde W}^v} - {\Phi ^v} > }  \\
 \end{array}
\end{equation}
\end{small}

\vspace{-0.2cm}
The iteration procedure to update the parameters is carried on until converges. Specifically, the variables $W$, $\tilde{W}$, $\varTheta$, $A$, $\varPhi$ and $\hat{W}$ are updated as follows, where $\tilde{W_{(i)}}$, $A_{(i)}$, $\varPhi_{(i)}$ are the values at $i$-th  iteration respectively.

\textbf{Update $\tilde{W}_{(i)}$:} Fix $W_{(i)}$ $\hat{W}_{(i)}$, $\varPhi_{(i)}$,  $\varTheta_{(i)}$, the function with respect to $\tilde{W}_{(i)}$ is:
\vspace{-0.12cm}
\begin{small}
\begin{equation}
\begin{array}{l}
\mathop {\min }\limits_{{{\tilde W}^v}} \sum\nolimits_v {\sum\nolimits_{v'} {||{A^v}^T{{\tilde W}^v} - {A^{v'}}^T{{\tilde W}^{v'}}||_F^2} }  + \sum\nolimits_v {u||{{\hat W}^v} - {{\tilde W}^v} - {\Phi ^v}||_F^2 +  < {\lambda _2},{{\hat W}^v} - {{\tilde W}^v} - {\Phi ^v} > }
\\
\vspace{-0.12cm}

 \end{array}
\end{equation}
\end{small}
\vspace{-0.12cm}

We can also solve it by gradient descent.

\textbf{Update $A_{(i)}$:} Fix $\tilde{W}_{(i)}$, the function with respect to $A_{(i)}$ is written as:

\begin{small}
\begin{equation}
\begin{array}{l}
\mathop {\min }\limits_{{A^v}} \sum\nolimits_v {\sum\nolimits_{v'} {||{A^v}^T{{\tilde W}^v} - {A^{v'}}^T{{\tilde W}^{v'}}||_F^2} }

 \end{array}
\end{equation}
\end{small}

\vspace{-0.12cm}

As is with \cite{wen2013feasible}, we can use matrix orthogonalization technique.  Then $A^v$ can be obtained by just setting derivation of $A$ regarding to Equation (9) to zero.  And finally, the $k$ smallest eigenvectors are selected to recover shared space ${A^v_{(i)}}^T$.

\textbf{Update $\varPhi_{(i)}$:} Fix $W_{(i)}$ and $\tilde{W}_{(i)}$, the function with respect to $\varPhi_{(i)}$ is written as:

\begin{equation}
\begin{array}{l}
\mathop {\min }\limits_{{\Phi ^v}} {\rho _3}||{\Phi ^v}|{|_{2,1}} + \sum\nolimits_v {u||{{\hat W}^v} - {{\tilde W}^v} - {\Phi ^v}||_F^2 +  < {\lambda _2},{{\hat W}^v} - {{\tilde W}^v} - {\Phi ^v} > }
 \end{array}
\end{equation}

As the same with updating $\varTheta$, it can be verified as a standard least absolute shrinkage and selection operator (LASSO) estimator.

\subsection{Multiple View Shaker Detection}
\vspace{-0.25cm}

The Markovian nature of the $RIN$ model enables us to perform predicative ``what-if''
analysis using
simulation to detect shakers.   That is, to measure the impact of an entity $u$, one may arbitrarily inject a small deviation ($\textit{e.g.}$, 1) and then gauge the
consequences of the injected conditions by predicting the future changes of the whole system.

The changes for entity $u$ after a small injection ($\textit{e.g.}$, 1) can be formalized as $||\hat{{W_u}}^r|{|_2}$, where $r$ is the  time interval\footnote{The parameter $r$ is the influence time duration, and it is application dependent.}. We define accumulated influence matrix by information cascades  within $r$ steps as $\sum\nolimits_{r} {{\hat{W}}^r}$.  As shown in the Phase 2 in Figure 3, the \textbf{ multi-view shakers} can be identified as

\vspace{-0.0867852cm}

\begin{equation}
U= \{ u:max\{f_u\} \} \quad s.t. \quad f_u=|| E_u|{|_2}, \quad E=\sum\nolimits_{r} {{\hat{W}}^r},
\end{equation}

\vspace{-0.0867852cm}

where $f_u$ represents the strength of the influence of $u$.   We explain why we define and utilize accumulated influence matrix with a typical example\footnote{https://www.ted.com/watch/ted-institute/ted-bcg/min-zhu-interconnectivity-the-new-structure-of-the-world-economy}  as follows: according to IMF survey, when GDP of USA falls by 1 percent, the direct impact on France is about 0.1 percent. However, the impact changes of United States can also spread to other countries around the world, which would amplify to 0.4 percentage to France by indirect impact of the United States in turn. This phenomenon is also called ``spill-over effect'' in economics. Note that though diffusion decay over the time period needs to be taken into account, it is beyond the scope of this paper and has not discussed.


%
%
%
%
%
%
%
%
%
%

\vspace{-0.43cm}

\section{Experiments}

\vspace{-0.13cm}

In this section, we present various experiments to evaluate the effectiveness of the proposed approach.  The program is implemented by MATLAB, and all experiments are running in Intel-core CPU Windows 7 and 32GB memory professional system.

\vspace{-0.1054312cm}

\subsection{Experiment Settings}

\vspace{-0.1054312cm}

We describe the dataset, baseline methods and model evaluation in detail in this part.

\textbf{Data Description:} To evaluate the performance, the following two dataset are tested.

\begin{itemize}

\vspace{-0.2432cm}

\item[1)] \textbf{The stock market dataset} contains reinstated closing prices and trading volumes  from January 2nd in the year 2008 to April 9th in the year 2016 on 1109 stocks in  Chinese stock market. Others are ignored for the sake of incomplete information, such as new entrants, $\textit{etc.}$

\item[2)] \textbf{The synthetic dataset} contains 17,744 entities within the time interval of 30. It is used to test the model scalability.

\end{itemize}

\vspace{-0.10653cm}

\vspace{-0.243cm}

\textbf{Baselines:} Note that we have proposed a new problem and no direct methods are specifically designed to solve our proposed problem due to the totally new settings - short-term characteristics, we choose classic methods that can provide intermediate results representing as real-valued score indicator for comparison. We compare the proposed model $RIN$ with the following baselines:
 \begin{itemize}

\item \textbf{sshaker:}   The sshaker (\cite{shi2011discovering1}) adapts polynomial mapping to infer influence relations between entities. The model utilizes basic function transformations (such as scaling, shifting, $\textit{etc.}$) to match every two time series, and encodes influence relation with shifting similarity degree.

\item \textbf{dshaker:} The dshaker algorithm (\cite{shi2013dynamic}) is another baseline. The linear trace minimization is used to estimate the influence relation. As with $RIN$,  it assumes the activity of the entity is affected by other entities under the Markovian assumption. A  multi-objective optimization problem is formulated  to learn the influence relation.

\item \textbf{tensor:} The tensor based method  (\cite{Bahadori2014Fast}) is used in exploiting multi-view information. This method is not designed for the same purpose as $RIN$ model, but its parameters can be  taken as influence strength when treating history information as input.

\end{itemize}

Meanwhile, to well understand how identifying different outliers works, two simplified versions of our model are proposed: (1) \textbf{sRIN1:} the model without considering \textbf{Outlier \uppercase\expandafter{\romannumeral1}}; (2) \textbf{sRIN2:} the model without considering \textbf{Outlier \uppercase\expandafter{\romannumeral2}}.

\textbf{Model Evaluation:} We evaluate the performance from synthetic data and  real data.  As for influence relation, we evaluate it on relation recovery from synthetic dataset directly. As for shaker detection,  we propose to evaluate the results based on their predictive ability in forecasting the whole market change. Meanwhile, model scalability and parameter sensitivity are also discussed. Furthermore, we show a case study of the shaker based algorithmic trading design and performance.

%
%
%
%
%
%
%

\subsection{Shaker Discovery on Real Dataset:}

%
%
%
%
%

\textbf{Shaker Based Market Trend Prediction:} Although the inference results in Table 4 make sense, they cannot be used to evaluate different methods quantitatively. Indirectly, we turn to discovered shaker based market trend prediction accuracy for evaluation.  Given the
precedent trend of shakers, we can predict the preceding  trend value of the whole system. More concretely, we assume the whole market change is mainly driven by these $U$ shakers, and the movement of the shakers can diffuse in the following $r$ days. That is to say, the whole market change can be monitored by a weighted combination of the evolution of these shakers. Mathematically, we  \textbf{monitor} the future market trend ${{\mathbf{m}_{t+r}}}$ based on the current trend ${\mathbf{m}_{t}}$ as
${{\mathbf{m}_{t+r}}} = {{{\mathbf{m}_{t}}}}\sum\limits_{u = 1}^{|U|} \frac{{{\mathbf{Y}}(t, u)}}{{{\mathbf{Y}}(t -1, u)}} {f_u}/\sum\limits_{u = 1}^{|U|} {{f_u}}$. The trend (price) of  a shaker $u$ is represented as ${{\mathbf{Y}}(t, u)}$, the movement ratio of shaker $u$  at time $(t - 1, t]$  as  $\frac{{{\mathbf{Y}}(t, u)}}{{{\mathbf{Y}}(t-1, u)}} $ , and  the influence strength of shaker $u$  as ${f_{u}}$  in Equation (11).

%
%
%
%
%
%
%

The trends of indexes $SH000300$ and  $SH000001$ in $SSE$ (Shanghai Stock Exchange, China) are chosen as the \textbf{ground truth}, represented as $\mathop {{\mathbf{m}_t}}\limits^ \to $. The rational lies in that these two indexes are  designed to measure  the overall trend of the market, which is also the meaning of the shakers. We \textbf{evaluate} the similarity between our predicted market change and ground truth  with the mean absolute error metrics as
$sMAPE = \frac{1}{{|T|}}\sum\limits_{t = 1}^{|T|} {\frac{{|{\mathbf{m}_t} - \mathop {{\mathbf{m}_t}}\limits^ \to  |}}{{({\mathbf{m}_t} + \mathop {{\mathbf{m}_t}}\limits^ \to  )/2}}}$. The smaller $sMAPE$ value is, the more effectiveness of the discovered shakers are in predicting changes. The initial values of both indexes $SH000300$ and $SH000001$ are normalized to 100 for ease of computation.

\subsection{Algorithmic Trading Toy Example}

%
%
%
%
%
%
%
%

\vspace{-0.01504106432cm}
\begin{figure}[b]
\vspace{-0.013242cm}
\centering

 \subfigure[Prediction Error  ($000300$)]{
      \includegraphics[width=0.46532102238302831253\textwidth,height=.245048637026804246248900251\textwidth]{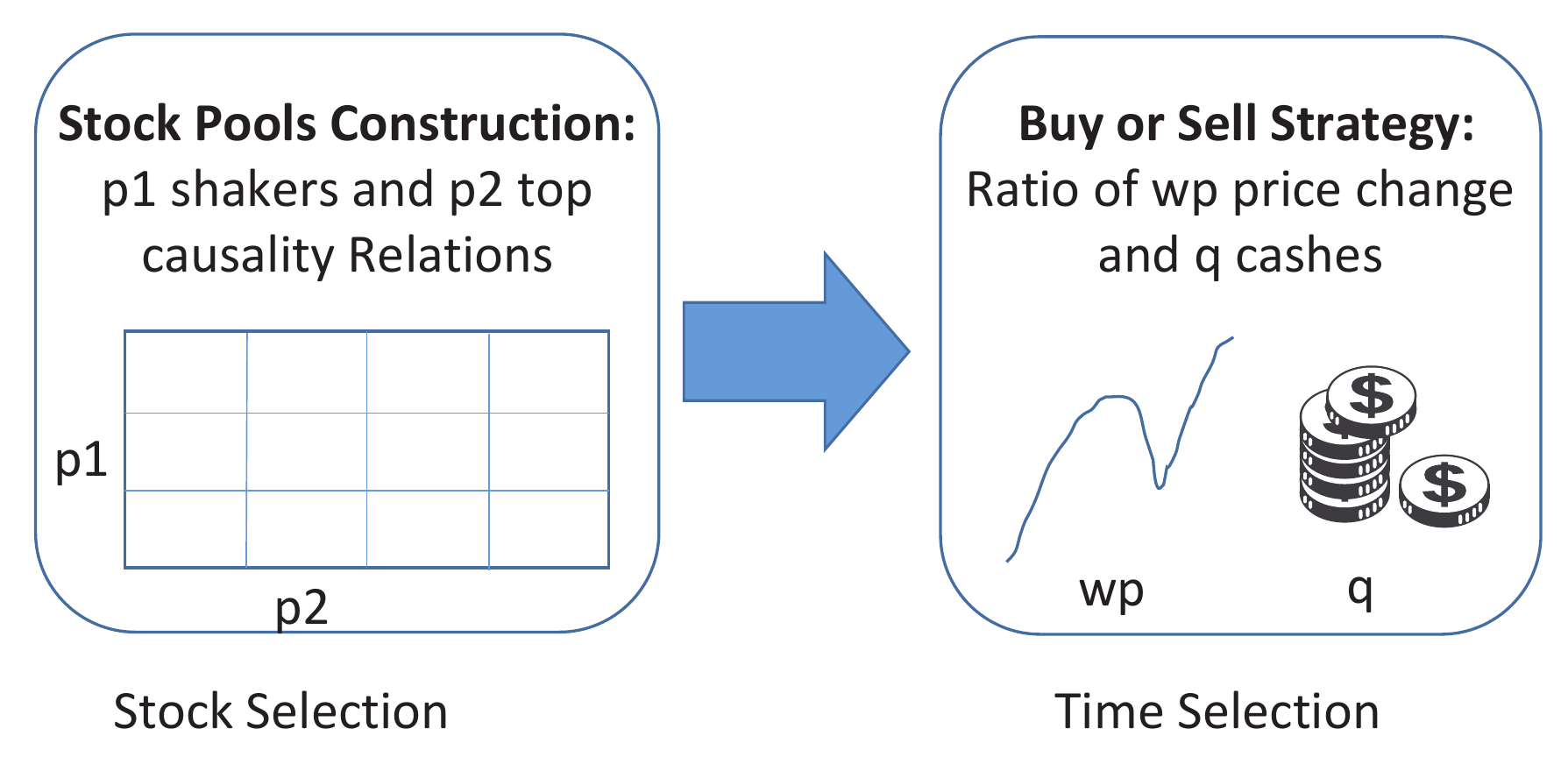}

  }
  \subfigure[Prediction Error ($000001$)]{
      \includegraphics[width=0.420321022389302831253\textwidth,height=.24853149637026804246248900251\textwidth]{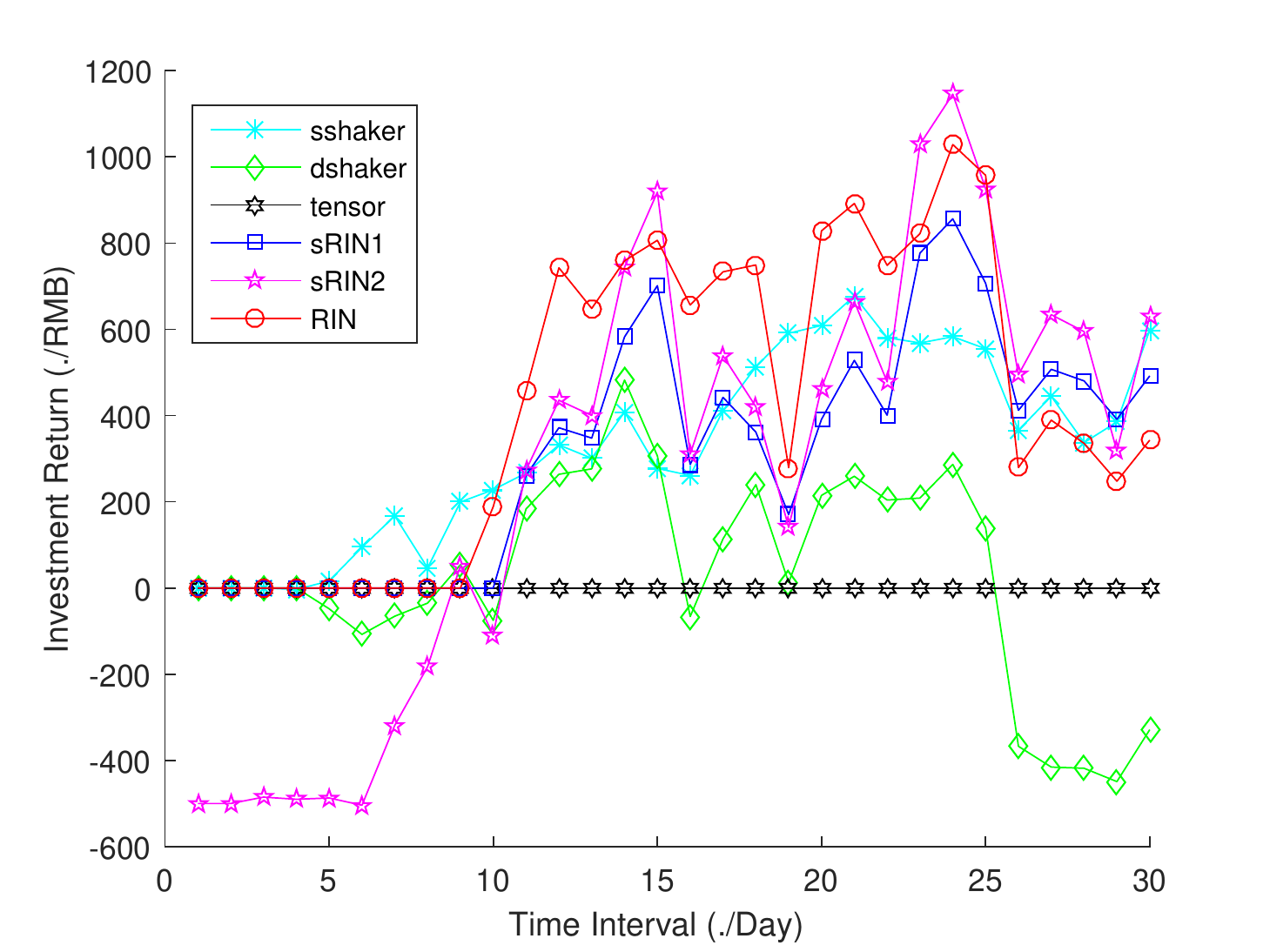}

  }

\vspace{-0.15432cm}
\caption{Toy Example of Shaker Based Algorithmic Trading}
\vspace{-0.5432cm}

\end{figure}

\vspace{-0.016405432cm}

 We compare the effectiveness of the learnt influence relation strength indirectly by  algorithmic trading results. As in Figure 6(a), the trading strategy  based on learnt influence relation and shakers consists of two parts: (1) stock selection \textbf{($Screening(p_1,p_2)$)}, we pick the top $p_1$ shakers  and  accumulate top $p_2$ influence relation for each shaker , and thus $p_1\times p_2$ candidate pairs are chosen; (2) time selection \textbf{($Timing(wp,q))$}, when the price of the shaker increases by $wp$, we ought to buy the stocks in the shaker's corresponding pools with $q$ portion of  the remaining cash, and otherwise we ought to sell the stocks.

We illustrate with \textbf{a toy example of the strategy performance} on real market data in 30 trading days time interval from October 23rd to December 3rd in Year 2015 (other time intervals for back test are also flexible). To begin, we have $RMB$ $10,000$ (approximately $USD$ $1,630$) and suppose that transaction cost is 0.  We compare the daily earning performances based on the shakers and  influence relation strength discovered by different methods.


\vspace{-0.35432cm}

\section{ Related Work}
\vspace{-0.13231cm}

We investigate the related work from the following four perspectives.

\textbf{Influence Inference From Time Series} Influence inference has been a hot research topic in recent years. Based on time series data,  a collection of techniques to discover and model influence relations have been proposed, such as \cite{granger1969investigating,liu2010learning}. The models to  detect shakers in economic by analyzing the behavior of time series data are proposed in (\cite{shi2011discovering1,shi2013dynamic}).   Extensions can be from linear relationship to non-linear relationship modeling (\cite{matsubara2016ecosystem}). By contrast, we extend influence inference  with outlier detection. There is some work (such as \cite{JalaliS12,Yang2013Mixture,Wang2013Sparse}, etc) that also consider latent factors (outliers) in influence inference, but none of them  have emphasised on short-term settings and identified two types of outliers thoroughly, especially in financial market.

\vspace{-0.0510231cm}

\textbf{Multi-view Learning} Recently, many prediction tasks aims to learn from instances that
have multiple views in different feature spaces (\cite{Li2017Multi}). Meanwhile, there are many multi-view multi-task learning
methods (\cite{Jin2013Shared,song2015interest}), which can make a good use of
the information contained in different tasks and views. As with influence inference, \cite{E2014LEARNING} proposes to infer influence relation between entities by utilizing  multi-source information.  To the best of our knowledge, limited work has been applied to influence inference from multiple view time series jointly.

\vspace{-0.0510231cm}

\textbf{Outlier Detection} Outlier detections have been well studied in recent years. \cite{Chen2011Integrating} proposes to capture the task relationships using a low-rank structure, and simultaneously
identifies the outlier tasks using a group-sparse structure. When interacted with multiple source/view  settings, \cite{gao2011spectral1} proposes to identify objects that have
inconsistent behavior  across multiple sources. \cite{Fakhraei2015Collective} aims to identify spammers in evolving multi-relational social networks, and \cite{zhao2015dual} attempts to discover different types of outlier in multi-view data.  As an important supplement and extension, we propose to infer robust influence relations from multi-view time series along with identifying different types of outliers. To the best of our knowledge, limited work has considered influence analysis and outlier detection tasks jointly.

\vspace{-0.0510231cm}

\textbf{Financial Data Analytic} Financial market analytic has attracted a great deal
of attention for years. From influence inference perspective,   \cite{bae2003new} proposes an approach to measure financial contagion. As with multi-view information, \cite{li2015tensor} proposes
to investigate the joint impact of different information
sources on stock markets with tensor based method.  \cite{Borboudakis2016Towards} attempts to propose a robust causal discovery method for business applications by integrating many  domain features. Above all, it's the first time to exploit influence relation discovery from time series in financial market joint with outliers detection under multi-view settings.

\vspace{-0.32043231cm}

\section{Conclusion and Future Work}

\vspace{-0.1043231cm}

This paper has described and studied a new problem - multi-view shaker detection. The problem is motivated from real-world applications under very short-term setting.  This is a novel application problem that has not been touched, which yet can accommodate a variety of application scenarios.  Aiming at it, we propose a general framework and a principled approach to detect shakers together with outlier awareness. The problem  is interesting and meaningful, and the proposed approach is also inviting.  Experimental results on both synthetic and real-world data show that the proposed model is effective. One potential work  is to consider nonlinear relationship.

\vspace{-0.6432cm}

\bibliographystyle{splncs03}
\bibliography{shaker-dasfaa}

\end{document}